\documentclass[11pt]{article}

\usepackage{amsmath} 
\usepackage{amsfonts} 
\usepackage{amssymb}             

\usepackage[dvips]{graphicx} 

\begin{document}

\newcommand{\sw}{\!\not \!}
\newcommand{\del}{\partial}
\newcommand{\noi}{\noindent}
\newcommand{\idd}{\mathbb{I}}
\newcommand{\nc}{\hat{n}_{i}}
\newcommand{\mc}{\hat{m}_{i}}
\newcommand{\mn}{\hat{\mbox{\boldmath$m$}}\cdot\hat{\mbox{\boldmath$n$}}}
\newcommand{\tg}{\mbox{tg}\,}
\newcommand{\qm}{\mbox{\boldmath$q$}}
\newcommand{\um}{m_{u}}
\newcommand{\cm}{m_{c}}
\newcommand{\tm}{m_{t}}
\newcommand{\dm}{m_{d}}
\newcommand{\sm}{m_{s}}
\newcommand{\bm}{m_{b}}
\newcommand{\elm}{m_{e}}
\newcommand{\mum}{m_{\mu}}
\newcommand{\tam}{m_{\tau}}
\newcommand{\nelm}{m_{\nu_e}}
\newcommand{\nmum}{m_{\nu_{\mu}}}
\newcommand{\ntam}{m_{\nu_{\tau}}}
\newcommand{\ru}{r_u}
\newcommand{\rd}{r_d}

\newcommand{\Det}{\mbox{det}\,}
\newcommand{\Tr}{\mbox{tr}\,}

\title{
\vspace{-2.25cm}
\parbox{\textwidth}{\small
    \begin{center}
      \hfill
      \parbox{6cm}{
        \begin{flushright}
          UCL-IPT-02-05
        \end{flushright}
        }
    \end{center}
    }\\[0.5cm]
\huge Finite Mixing Angles in the Standard Model}

\date{\vspace{-1.5cm}}

\author{M. Buysse\thanks{E-mail: buysse@fyma.ucl.ac.be} \\
{\it Institut de Physique Th{\'e}orique }\\
{\it Universit{\'e} Catholique de Louvain}\\
{\it 2, Chemin du Cyclotron} \\
{\it B-1348 Louvain-La-Neuve, Belgium}\\
}

\maketitle

\begin{abstract}
We establish the existence of {\it one-loop finite relations}
between the Cabibbo angle and the quark mass ratios, in the Standard
Model with one Higgs doublet and two quark generations.
Assuming a simple quark-lepton universality, we use the recent
SNO results to predict the two-flavour mass spectrum
of the neutrinos. 
\end{abstract}

\normalsize

\section{Introduction}

Searching for relations between the apparently free parameters of
the Standard Model (SM) of elementary particles ranks among the
most popular unification attempts. Most of those parameters find their origin in the
Yukawa sector of the theory. In particular, the six
quark masses and the three Cabibbo-like mixing angles exhibit a strongly
hierarchical pattern among families. Furthermore, recent experiments
tend to confirm the existence of neutrino masses and therefore
of new mixing angles in the lepton sector \cite{Bahcall:2002}. For decades one has tried to
constrain the SM in order to obtain relations
between those apparently free parameters. 
Until now, most attempts consisted in enlarging the symmetry
group of the SM by
adding a horizontal component to it 
\cite{Wilczek:1977}. 
The horizontal symmetry imposes constraints on the structure of the
Yukawa couplings. After spontaneous breakdown of the symmetry, the
fermion mass matrices that are generated still bear the stamp of those
constraints and through bidiagonalization, they give rise to relations
between mass ratios and mixing angles. Such an implementation
guarantees that the relations survive to renormalization; they are
called {\it natural} \cite{Weinberg:1974}. 
However one soon realized \cite{Barbieri:1978} that it could not be achieved without
extending the particle content, namely by considering models with more than
one Higgs doublet, thereby increasing
the number of couplings...

In this paper we present an alternative way to envisage naturalness in
the Yukawa sector. We assume the existence of a tree-level relation
between some apparently free parameters, such that it is not spoiled
by divergent one-loop radiative corrections. The condition is
necessary to get natural relations (it is not sufficient as long as the
relation does not
hold at all orders in perturbation theory). Moreover, instead of one-loop naturalness, 
we talk about one-loop finiteness, since the obtained result is in
principle not correlated with the presence of an extra symmetry group in
the SM. 

\vspace{10mm}

A first section is devoted to the presentation of a situation (evoked
in \cite{Branco:1983}) where the use of a simple
one-loop argument happens to be quite powerful. We show how such a kind of argument, applied
to the weak mixing angle $\theta_{W}$, within
the SM, may give an insight of some higher scale {\it vertical}
symmetry, and suggest the existence, at that scale, of a Grand Unified
Theory (GUT).

In the second section we focus on the Cabibbo mixing angle $\theta_{C}$ \cite{Cabibbo:1963} 
in a SM with one Higgs doublet and two fermion generations. Again we
make use of a one-loop argument to show that it is
impossible to determine a non-trivial Cabibbo angle as a fixed,
calculable, parameter. The argument confirms a well-known result about {\it horizontal}
symmetries with a single Higgs doublet \cite{Barbieri:1978}. 

We then consider a larger set of {\it a priori} free parameters
in the hadronic Yukawa sector of the SM with one Higgs doublet and two fermion
generations, and we establish a method to find out the one-loop finite
relations between them. We show that there is indeed an infinite set of
one-loop finite relations (expressing the Cabibbo angle $\theta_{C}$ as a
function of the quark mass ratios $\um/\cm$ and  $\dm/\sm$) in the
single-Higgs-doublet SM. We conclude that those
relations cannot originate from any additionnal horizontal symmetry. 

Assuming the
existence of a purely Dirac mass term for the neutrinos, we obtain a
second infinite set of one-loop finite relations in which the lepton
mixing angle $\theta_L$ is found to be a non-trivial function of the lepton mass
ratios $\nelm/\nmum$ and $\elm/\mum$. 

Finally, using the Large Mixing Angle (LMA) solution favoured by
the new results of the SNO collaboration \cite{Ahmad:2002a},
we predict
the two-flavour mass spectrum of the neutrinos from a simple
quark-lepton universality.

\section{The Weak Angle}

The weak angle $\theta_{W}$ defines the physical $A$ and $Z^0$ neutral
gauge bosons of the tree-level $SU(2) \otimes U(1)$
Standard Model. This angle is in principle renormalized by the 
vacuum polarization diagrams that mix $A$ and $Z^{0}$. 

Now imagine some new theory, beyond the SM, in which the weak angle is no
longer a free parameter. In the absence of counterterm, the radiative
corrections to it would then be finite. In
particular, at the one-loop level, the fermionic contribution 
\begin{equation}
\parbox{30mm}{\includegraphics[width=30mm,keepaspectratio]{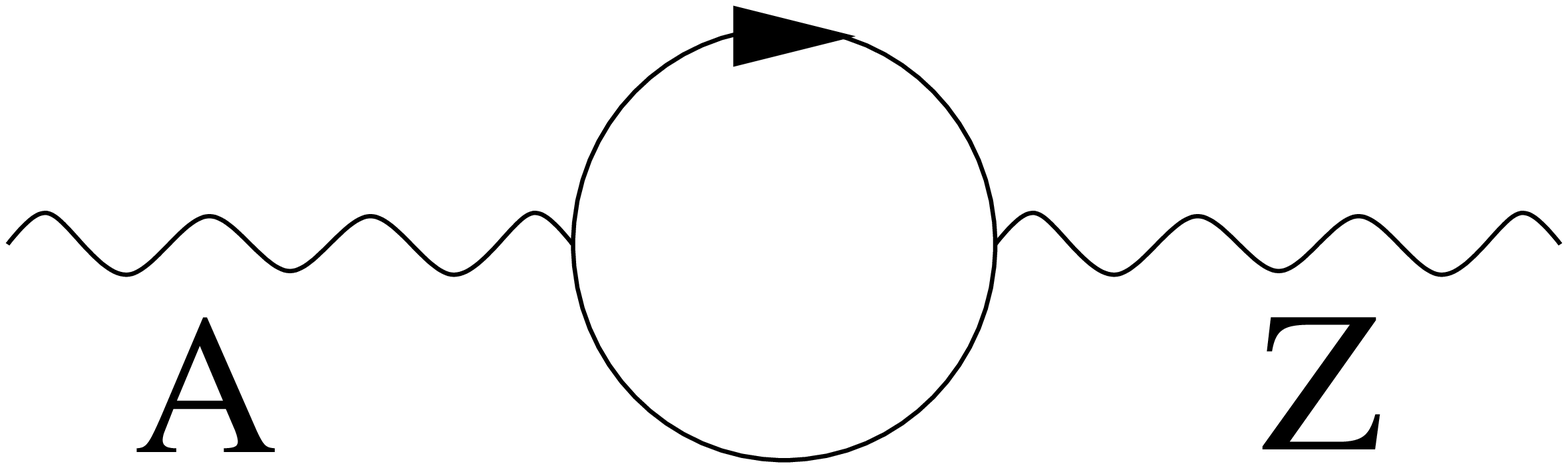}}
\label{pola}
\end{equation}

\noi would be finite. Therefore, if the matter content of the new
theory is identical to the one of the SM, the
divergent contributions of the diagram (\ref{pola}) vanish. At the one-loop
level, the divergent part
of the corrections to $\theta_{W}$ proportional to the number $n$ of
fermion generations reads: 
\begin{equation}
\delta \, \theta_{W}\,
\sim
\, n \, \left (1-\frac{8}{3} \sin ^{2} \theta_{W}\right )
\,
\ln \frac{\Lambda^2}{\mu^2}
\label{condifaible}
\end{equation}

\noi where $\Lambda$ is a cut-off\footnote{We did not write down the
  quadratically divergent term since it can be made irrelevant in
  other regularization schemes. However, including this term in
  equation (\ref{condifaible}) does not affect the form of the overall
  coefficient.}
and $\mu$ an arbitrary energy scale. 
Hence, putting those divergent contributions to zero amounts to impose
the non-trivial tree-level relation
\begin{equation*}
\sin ^{2} \theta_{W}
= 
\frac{3}{8}\quad .
\label{sintgu}
\end{equation*}

\noi Such a constraint on the weak angle does not match the measured value 
\begin{equation*}
\sin ^{2} \theta^{exp}_{W} \simeq  0.23
\end{equation*}

\noi and, as such, should be thrown away. But our basic assumption was
about the existence of some higher scale theory -- or some higher
scale symmetry which would
determine $\sin ^{2} \theta_{W}$ at that very scale. From this point
of view, one notices that $\frac{3}{8}$ is precisely the value of
$\sin ^{2} \theta_{W}$ predicted at the $SU(5)$ GUT scale. So that one can expect
$\sin ^{2} \theta_{W}$ to run from its $SU(5)$ value to its
SM value (partially due to the finite part of the diagram (\ref{pola})
if the fermion
masses are taken to be non-degenerate within each generation). 

Furthermore, equation (\ref{condifaible}) is also true, independently, for two subsets of
fermions running in the loop of the diagram (\ref{pola}), namely the left-handed
electron, the left-handed neutrino\footnote{The neutrino has no charge
  and does not couple to the photon, but we mention it since it is
  associated to the electron through the $SU(2)$ doublet structure. }
and the three coloured right-handed
$down$ quarks on the one hand, and the remaining ten SM fermions of
the first generation on the other
hand. This simply tells us the minimal way one could distribute the SM
fermions in irreducible representations of the higher scale
symmetry group, with well-defined dimensions: 
\begin{equation*}
\nu_{L}, e_{L}, d_{R}^{r}, d_{R}^{b}, d_{R}^{g}  \to \mbox{\boldmath$5$}
\qquad\mbox{and}\qquad
\mbox{others}  \to \mbox{\boldmath$10$}\quad .
\end{equation*}

A one-loop argument within the SM provides therefore some information about
hypothetical vertical gauge symmetries (GUT's). It is
then tempting to adopt the same procedure for the Cabibbo mixing
angle. 

\section{The Cabibbo Angle}

In the two-fermion-generation SM, the Cabibbo angle $\theta_{C}$ mixes the isospin
eigenstates quarks, to end up with the mass eigenstates ones. At the
one-loop level, it is
radiatively corrected by a combination of self-energy diagrams of the
following type: 
\begin{equation}
\parbox{30mm}{\includegraphics[width=30mm,keepaspectratio]{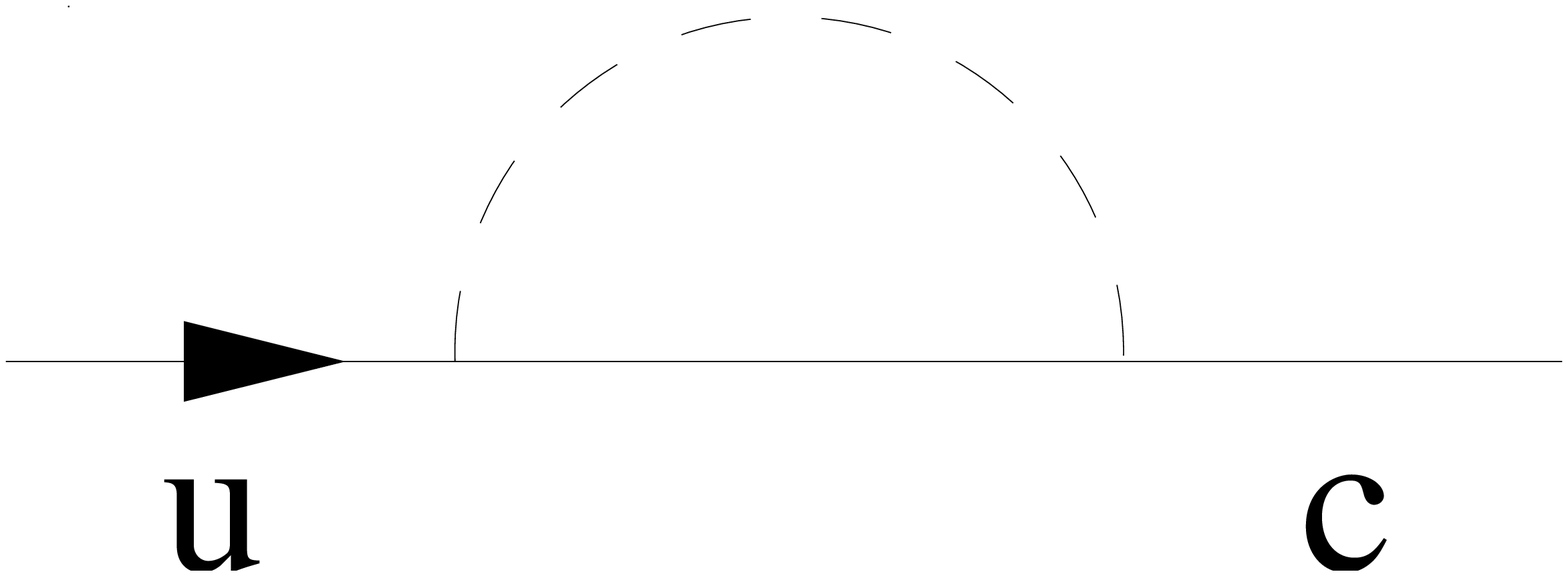}}
\qquad\qquad\qquad\qquad
\parbox{30mm}{\includegraphics[width=30mm,keepaspectratio]{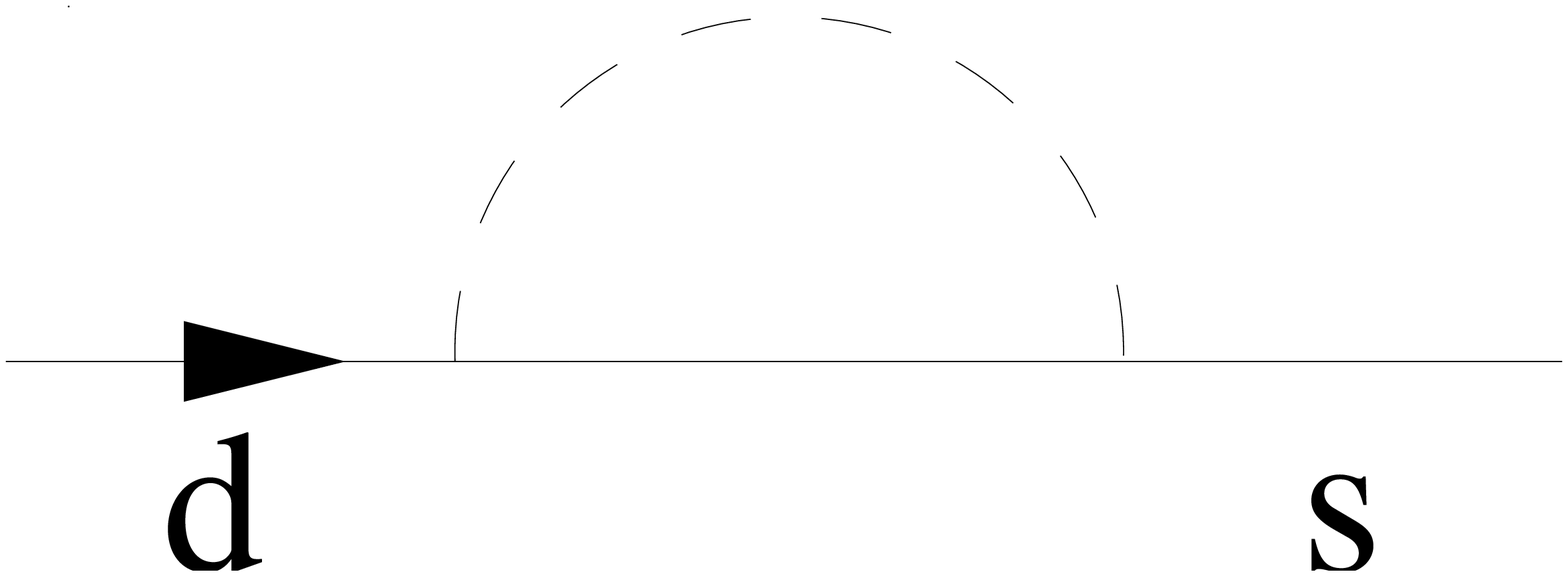}}
\label{self}\quad .
\end{equation}

Now, one may ask again the question: is there a value of $\theta_{C}$ such that
the radiative corrections to it are finite? Let us put the
divergent contributions of the right combination of the diagrams (\ref{self}) to
zero, to obtain a constraint on the Cabibbo angle. At the one-loop
level, in a single-Higgs-doublet model, 
the divergent part of the corrections to $\theta_C$
reads\footnote{The exact expression, given by the first equation
  of (\ref{varpara}), is computed in the Appendix. }: 
\begin{equation}
\delta \theta_C
\,  \sim \, \sin \theta_{C} \cos \theta_{C} 
\,
\ln \frac{\Lambda^2}{\mu^2}
\label{condicabi}
\end{equation}

\noi where $\Lambda$ is a cut-off and $\mu$ an arbitrary energy scale. 
Hence, putting those divergent contributions to zero amounts to impose
the constraint
\begin{equation*}
\sin \theta_{C}
= 
0
\qquad\qquad\mbox{or}\qquad\qquad
\cos \theta_{C}
=
0
\end{equation*}

\noi that is a trivial
mixing. This is not a satisfactory result since it does not
correspond to the SM value 
\begin{equation*}
\sin  \theta^{exp}_{C} \simeq  0.22 \quad .
\end{equation*}

\noi Moreover a trivial mixing angle cannot be associated to any
higher scale symmetry, since
it could not run to the non-zero SM value (with a trivial mixing,
even the finite part of the diagrams (\ref{self}) vanishes, so that a non-trivial mixing cannot
originate from a tree-level trivial mixing\footnote{A tree-level trivial
  mixing implies the existence of a $U(1)$ symmetry for each
  generation, which guarantees the absence of mixing at all orders in
  perturbation theory. }). 

This simple one-loop argument agrees with a well-known theorem
\cite{Barbieri:1978}, which states that horizontal
symmetries cannot lead to the non-trivial determination of the Cabibbo
angle in the single-Higgs-doublet SM.

\section{The Cabibbo Angle and the Quark Mass Ratios}

One may think that the strict vanishing of the divergent corrections
to $\theta_C$ is too strong a requirement. Inspired by the numerical
success of $\tan \theta_C =\sqrt{m_d/m_s}$ \cite{Gatto:1968}, one can
indeed relax this requirement and
try to find a similar relation such that the divergent corrections to
$\theta_{C}$ exactly compensate the corrections to the mass ratios. We
know that this kind of relation cannot be obtained via the use of
horizontal symmetries in the single-Higgs-doublet SM \cite{Barbieri:1978}. However, the
existence of one-loop
finite relations is not necessarily related to the presence of any horizontal symmetry.

We consider the two-fermion-generation SM in its minimal realization, namely built up with one
single Higgs doublet. The hadronic Yukawa sector, which we are
interested in, contains three useful free parameters: the Cabibbo angle
$\theta_{C}$ and the two mass ratios $\ru =\um/\cm$ and  $\rd
=\dm/\sm$. 
We assume the {\it a priori}
existence of a relation between the Cabibbo angle and the mass
ratios, 
whose general form reads: 
\begin{equation}
F(\theta_{C})=G(\ru , \rd)
\label{relnat}
\end{equation}

\noi so that 
\begin{equation}
F(\theta_{C} + \delta \, \theta_{C}) 
= G(\ru + \delta \, \ru , \rd + \delta \, \rd)
\label{propnat}
\end{equation}

\noi where $\delta \, \theta_{C}$, $\delta \, \ru $ and $\delta \, \rd$ are the divergent
parts of the radiative corrections to $\theta$, $\ru$ and $\rd$
respectively. Thus
\begin{equation*}
F_{\shortmid \theta_{C}}\, \delta \, \theta_{C} 
=
G_{\shortmid r_{u}}\, \delta \, \ru 
+ 
G_{\shortmid r_{d}} \, \delta \, \rd 
\end{equation*}

\noi where $f_{\shortmid x}$ denotes the (partial) derivative of $f$
with respect to $x$. 
Computing the divergent part of the one-loop radiative corrections to the Yukawa couplings 
(see Appendix), one obtains: 
\begin{equation}
\begin{matrix}
\delta \, \theta_{C} 
 = 
\epsilon \left [
\frac{1+\ru ^{2}}{1-\ru ^{2}}(\dm ^{2} - \sm ^{2})
+ 
\frac{1+\rd ^{2}}{1-\rd ^{2}}(\um ^{2} - \cm ^{2})
\right ]
\sin \theta _{C}\cos \theta _{C}
\\
\\
\delta \, \ru
=
\epsilon \; \ru \;
[ (\dm ^{2} - \sm ^{2})
\cos 2 \theta_{C} 
- (\um ^{2} - \cm ^{2})
]
\label{varpara}
\\
\\
\delta \, \rd
= 
\epsilon \; \rd \;
[ (\um ^{2} - \cm ^{2})
\cos 2 \theta_{C} 
- (\dm ^{2} - \sm ^{2})
]
\end{matrix}
\end{equation}

\noi with $\epsilon = \frac{3}{4} \frac{2}{v^{2}}\left (\frac{1}{4\pi^{2}}\ln
  \frac{\Lambda^{2}}{\mu^{2}} \right )$, $\Lambda$ being a cut-off, $\mu$ an
arbitrary energy scale and $v$ the vacuum expectation value of the
scalar field. Introducing those expressions into equation (\ref{propnat})
splits it into two independent equations (since we are only interested in
relations involving the mass ratios $\ru$ and
$\rd$): 
\begin{equation}
 F_{\shortmid \theta_{C}}\sin \theta_{C} \cos \theta_{C} 
\,\frac{1+\ru ^{2}}{1-\ru ^{2}}
= 
G_{\shortmid r_{u}}\,\ru \cos 2 \theta_{C}
-
G_{\shortmid r_{d}}\,\rd 
\label{sys1}
\end{equation}
\vspace{-4mm}
\begin{equation}
 F_{\shortmid \theta_{C}}\sin \theta_{C} \cos \theta _{C}
\,\frac{1+\rd ^{2}}{1-\rd ^{2}}
= 
G_{\shortmid r_{d}}\,\rd \cos 2 \theta_{C}
- 
G_{\shortmid r_{u}}\,\ru 
\label{sys2}
\end{equation}

\noi which turn out to be compatible if and only if 
\begin{equation}
\cos 2\theta_{C} 
= 
\frac
{
\frac{1-\ru ^{2}}{1+\ru ^{2}}\,
\rd\,
G_{\shortmid r_{d}}
-
\frac{1-\rd ^{2}}{1+\rd ^{2}}\,
\ru\,
G_{\shortmid r_{u}}
}
{
\frac{1-\ru ^{2}}{1+\ru ^{2}}\,
\ru\,
G_{\shortmid r_{u}}
-
\frac{1-\rd ^{2}}{1+\rd ^{2}}\,
\rd\,
G_{\shortmid r_{d}}
}
\label{identif}\quad .
\end{equation}

\noi Now this {\it must} be the relation
(\ref{relnat}) whose
existence has been assumed, i.e.\footnote{One checks that the
arbitrary character of those identifications will not
show itself in the expected solution. To prove it, we imagine
(\ref{eqangle}) would rather read $f(F(\theta_{C}))=\cos 2 \theta_{C}$. One should
then replace $G$ by $f(G)$ in the left-hand side of
(\ref{eqmasses}). But the right-hand side of it is
invariant under $G\mapsto f(G)$.  Namely, one can solve
(\ref{eqmasses}) with respect to the variable $f(G)$ which we eventually
identify to $f(F(\theta_{C}))=\cos 2 \theta_{C}$.  }
\begin{equation}
F(\theta_{C})=\cos 2 \theta_{C}
\label{eqangle}
\end{equation}

\noi and
\begin{equation}
G(\ru ,\rd)
= 
\frac
{
\frac{1-\ru ^{2}}{1+\ru ^{2}}\,
\rd\,
G_{\shortmid r_{d}}
-
\frac{1-\rd ^{2}}{1+\rd ^{2}}\,
\ru\,
G_{\shortmid r_{u}}
}
{
\frac{1-\ru ^{2}}{1+\ru ^{2}}\,
\ru\,
G_{\shortmid r_{u}}
-
\frac{1-\rd ^{2}}{1+\rd ^{2}}\,
\rd\,
G_{\shortmid r_{d}}
}
\label{eqmasses}\quad .
\end{equation}

\noi Exploiting the remaining information in (\ref{sys1}) and
(\ref{sys2}), and using (\ref{identif}) and (\ref{eqangle}),
yields 
\begin{equation}
\begin{cases}
\frac{1+\ru ^{2}}{1-\ru ^{2}} G
- \ru G_{\shortmid r_{u}} 
\!\!\!\!\!\! & + \frac{1+\rd ^{2}}{1-\rd ^{2}} 
 = 0 
\\
\frac{1+\rd ^{2}}{1-\rd ^{2}} G
- \rd G_{\shortmid r_{d}} 
\!\!\!\!\!\! & + \frac{1+\ru ^{2}}{1-\ru ^{2}} 
 = 0\quad .
\end{cases}
\label{sysedo}
\end{equation}

\noi This system -- from which one obviously recovers equation (\ref{eqmasses})
-- is integrable, and the general solution reads
\begin{equation*}
G(\ru , \rd)
= \frac{ - (1+\ru ^{2})(1+\rd ^{2}) + 2 \lambda_C \: \ru \rd}
{(1-\ru ^{2})(1-\rd  ^{2})}
\end{equation*}

\noi with $\lambda_C$ a flavour-blind integration constant! One concludes that, if a
relation of the kind suggested in (\ref{relnat}) exists, it necessarily belongs to
the following class:
\begin{equation}
\cos 2\theta_{C}
= \frac{\textstyle - (\um ^{2} + \cm ^{2}) (\dm^{2} + \sm ^{2}) 
+ 2\lambda_C \: \um\cm\dm\sm}
{\textstyle (\um ^{2} - \cm ^{2}) (\dm ^{2} - \sm ^{2})}
\label{sol}\quad .
\end{equation}

We have reached our goal to express the
Cabibbo angle as a function of the quark mass ratios. However, there
still is the unknown integration constant $\lambda_C$ which must be
greater or equal to $2$. One
checks that under the interchange $\um
\leftrightarrow \cm$ (or $\dm
\leftrightarrow \sm$), the Cabibbo angle of equation (\ref{sol}) moves
to its complementary, as expected. In
the realistic limit $\um << \cm$
and $\dm << \sm$, equation (\ref{sol})
implies
\begin{equation}
\cos ^2 \theta_C \simeq \lambda_C \frac{\um\dm}{\cm\sm}
\label{relapproq}
\end{equation}

\noi and requires therefore a large-valued $\lambda_C$ to get $\cos ^2
\theta_C $ close to its experimental value. 

\section{The Lepton Mixing Angle and Neutrino Masses}

The analysis we have conducted here can be applied to the leptons,
provided that the neutrinos are massive, their mass being of the Dirac
type exclusively. We thus consider the minimal extension of the SM that
allows for massive neutrinos. We add right-handed neutrinos to the
matter content
and maintain the lepton number conservation, thereby forbidding
Majorana mass terms. For the SM with two fermion generations and one
Higgs doublet, we perform the same
calculation\footnote{The calculation depends on the sole structure of
  the Yukawa couplings and is therefore identical.} as in the quark
sector. We conclude that any natural relation between the lepton mixing
angle $\theta_L$ and the lepton mass ratios must be of the following type:
\begin{equation}
\cos 2\theta_{L}
= \frac{\textstyle - (\nelm ^{2} + \nmum ^{2}) (\elm^{2} + \mum ^{2}) 
+ 2\lambda_{L}\:\nelm\nmum\elm\mum}
{\textstyle (\nelm ^{2} - \nmum ^{2}) (\elm ^{2} - \mum ^{2})}
\label{sollep}
\end{equation}

\noi $\lambda_{L}$ being again a flavour-blind integration constant greater or equal to
$2$. 
If $\nelm << \nmum$, then one has
the approximate relation 
\begin{equation}
\cos ^2 \theta_L \simeq \lambda_L \frac{\nelm\elm}{\nmum\mum}
\label{relapprol}
\end{equation}

\noi analogous to (\ref{relapproq}). If $\nelm >> \nmum$, we simply
interchange the neutrino masses in (\ref{relapprol}). 

\section{Quark-lepton Universality}

A one-loop calculation does not
determine neither $\lambda_{C}$ nor $\lambda_{L}$. These dimensionless
parameters could {\it a priori} take any real
value greater than $2$. But their flavour independence
prompts us to assume a simple quark-lepton universality such that
\begin{equation*}
\lambda_{C}=\lambda_{L}\quad .
\end{equation*}

\noi Combining equations (\ref{sol}) and (\ref{sollep}) in terms of
the mass ratios $r_u$, $r_d$, $r_{\nu}\equiv
\nelm/\nmum$ and $r_{\ell}\equiv
\elm/\mum$, we end up now with a one-loop finite relation
\begin{equation}
\frac{
(1 + r_{\nu} ^{2}) (1 + r_{\ell} ^{2}) 
+
(1 - r_{\nu} ^{2}) (1 - r_{\ell} ^{2}) 
\cos 2\theta_{L}
}
{
(1 + r_{u} ^{2}) (1 + r_{d} ^{2}) 
+
(1 - r_{u} ^{2}) (1 - r_{d} ^{2}) 
\cos 2\theta_{C}
}
=
\frac{r_{\nu}r_{\ell}}{r_u r_d}\quad .
\label{solmel}
\end{equation}
%

\noi involving only ``measurable'' quantities. In the limit 
$r_u << 1$, $r_d << 1$, $r_{\nu} << 1$ and $r_{\ell} << 1$, equation
(\ref{solmel}) yields
\begin{equation}
\frac{
\cos ^2\theta_{L}
}
{
\cos ^2\theta_{C}
}
\simeq
\frac{r_{\nu}r_{\ell}}{r_u r_d}\quad .
\label{solmelappro}
\end{equation}
%

\noi Introducing reasonable mass ratios \cite{Fusaoka:1998}, the Cabibbo mixing
angle and the LMA solution \cite{Ahmad:2002a}, one finds: 
\begin{equation*}
\frac{\nelm}{\nmum}\simeq\frac{1}{35}\quad .
\end{equation*}

\noi The mass splitting $\Delta
m^2=\nmum^2-\nelm^2$ associated to the LMA solution \cite{Ahmad:2002a}
implies
$\nmum\simeq 7\cdot 10^{-3} \mbox{eV}$ and $\nelm\simeq 2\cdot 10^{-4}
\mbox{eV}$.

\section{Conclusion}

We have derived two infinite sets of one-loop finite
relations from the Yukawa sector. Each of them expresses a
mixing angle as a function of fermion mass ratios. Determined by equations (\ref{sol})
and (\ref{sollep}), thoses sets are parametrized in terms of two
dimensionless constants $\lambda_C$ and $\lambda_L$. At the one-loop
level, one has no further theoretical argument to
constrain the value of those constants; and one cannot evade the difficulty by
asking one of the quark (or lepton) masses to vanish, since it would
then lead to a cosine
smaller than minus one\footnote{Consequently we may already conclude that it is not possible to naturally
set the mass of one single quark to zero, together with the
requirement of the existence of a natural relation between the Cabibbo
angle and the quark mass ratios. In this context, by excluding $\um = 0$ our result
rules out the natural vanishing of the QCD $T$-violating parameter
$\theta_{S}$.}. However, by identifying $\lambda_C$ and $\lambda_L$, which are flavour-blind,
we are then able to predict the two-flavour mass spectrum of the
neutrinos. 

Stating that finite relations potentially exist in a
single-Higgs-doublet model apparently contradicts previous results obtained in the
context of family symmetries \cite{Barbieri:1978}. But since we do not appeal to such kind
of symmetries, we do not expect our result to respect the conclusions
derived in their context. Namely, the one-loop finite relations
(\ref{sol}) or (\ref{sollep}) cannot be associated with the presence of any extra
horizontal symmetry. 
One should examine the validity of the
results at
higher loop level; then look for some possible ``determination principle'' of it outside or beyond the
SM -- just as the $SU(5)$ GUT determines the ``one-loop conjectured'' value of the weak angle
$\theta_{W}$ at the GUT scale. 

Whe should stress that the existence of non-trivial solutions to the system
(\ref{sysedo}) crucially relies on the expression of the divergent
one-loop radiative corrections to the Yukawa parameters, and would
definitively be invalidated if one modifies a single coefficient in those
corrections.

The extension of the present calculation to a three-generation
SM seems to be doomed to failure because of the complexity of the
one-loop radiative corrections to the Cabibbo-Kobayashi-Maskawa parameters
\cite{Cabibbo:1963, Kobayashi:1973}. Those
corrections are indeed too cumbersome to be manipulated and introduced
into a solvable partial differential equations system. We will however
expound, in a forthcoming paper, an alternative approach to
derive the one-loop finite relations in the $n$-generation SM.

\section*{Acknowledgments}
This work is supported by the {\it Fonds pour la Formation {\`a} la
  Recherche dans l'Industrie et dans l'Agriculture} (FRIA).  
We would like to thank J.-M. G{\'e}rard and J. Weyers for a critical
reading of the manuscript.

\section*{Appendix}

\appendix

\section{Introductory remark}

We start from a SM with one Higgs doublet and $n$ fermion
generations. 
The approach we put forward is based on the calculation of the
divergent one-loop radiative corrections to the quark mass matrices,
which exclusively involves self-energy and tadpole diagrams. More precisely, since we are
interested in natural relations between up-type quark mass ratios and 
down-type quark mass ratios on the one hand, and mixing angles on the
other hand, we will solely compute the divergent one-loop radiative
corrections to those specific parameters. This considerably simplifies
our task. One indeed notices that neither QED nor QCD, which are
flavour-blind, will bring in divergent contributions that would affect the
mixing angle or the mass ratios. The same argument holds for the diagrams involving the
transverse polarizations of the $Z^{0}$ and of the
$W^{\pm}$ vector bosons, as well as the tadpoles. 
To convince oneself, it is worth checking that the contribution of the
latter diagrams to the renormalization of the quark mass ratios reads
\begin{equation*}
\frac{m_{u}}{m_{c}} \longmapsto
\frac{(1+C_{\gamma}+C_{G}+C_{Z^{0}}+C_{W^{\pm}}+C_{T})m_{u}}
{(1+C_{\gamma}+C_{G}+C_{Z^{0}}+C_{W^{\pm}}+C_{T})m_{c}} 
=\frac{m_{u}}{m_{c}} 
\end{equation*}

\noi where $C_{\gamma}$, $C_{G}$, $C_{Z^{0}}$ and $C_{W^{\pm}}$ 
respectively originate from the interventions of the photon, the
gluons, the transverse $Z^{0}$ and the transverse $W^{\pm}$ in the up-type quark
self-energies, while $C_{T}$ originates from the tadpole diagrams. 
Those $C$'s are identical for the mass renormalization of
any up-type quark -- as far as the divergent part is concerned. The same
reasoning can be applied to the down-type quarks. For
the mixing angles, the proof is even more direct since none of those
diagrams leads to divergent non-diagonal correction to the tree-level
diagonal mass matrices. 
In other words, the only diagrams one has to consider are the quark self-energies
due to the exchange of the scalars (Higgs and would-be-Goldstone
bosons) -- for
the complete list of the relevant divergent diagrams, see figure
1. This is not astonishing since the scalars are the only fields
that know about the difference between the fermion families.

\section{Self-energies in the Yukawa sector}

After spontaneous breakdown of the symmetry, the Yukawa sector
Lagrangian for quarks reads:
\begin{equation*}
\begin{matrix}
\mathcal{L}_{Y} &
= & 
\bar u_{L}  \Gamma_{d} d_{R} \phi^{+} & 
+ & 
\bar d_{L}  \Gamma_{d} d_{R} \phi^{0} & 
+ & 
\bar d_{R}  \Gamma_{d}^{\dag} u_{L} \phi^{-} & 
+ & 
\bar d_{R}  \Gamma_{d}^{\dag} d_{L} \phi^{0\,\star} &
\\
& 
+ & 
\bar u_{L}  \Gamma_{u} u_{R} \phi^{0\,\star} & 
- & 
\bar d_{L}  \Gamma_{u} u_{R} \phi^{-} & 
+ & 
\bar u_{R}  \Gamma_{d}^{\dag} u_{L} \phi^{0} & 
- & 
\bar u_{R}  \Gamma_{u}^{\dag} d_{L} \phi^{+} &
\\
& 
+ & 
\bar d_{L}  M_{d} d_{R} & 
+ & 
\bar u_{L}  M_{u} u_{R} & 
+ & 
\bar d_{R}  M_{d}^{\dag} d_{L} & 
+ & 
\bar u_{R}  M_{d}^{\dag} u_{L} & .
\end{matrix}
\end{equation*}

\noi The Lagrangian fields and parameters are renormalized: 
\begin{equation*}
\begin{matrix}
u_{L,R} & 
\longmapsto & 
u'_{L,R} & 
= & 
(Z^{u}_{L,R})^{-\frac{1}{2}}u_{L,R} &\\ 
d_{L,R} & 
\longmapsto & 
d'_{L,R} & 
= & 
(Z^{d}_{L,R})^{-\frac{1}{2}}d_{L,R} &\\ 
M_{u} &
\longmapsto & 
M_{u}' & 
= & 
(Z^{u}_{L})^{\frac{1}{2}}
(Z_{M_{u}})^{-1}
M_{u}
(Z^{u}_{R})^{\frac{1}{2}}&\\
M_{u} &
\longmapsto & 
M_{d}' & 
= & 
(Z^{d}_{L})^{\frac{1}{2}}
(Z_{M_{d}})^{-1}
M_{d}
(Z^{d}_{R})^{\frac{1}{2}}&.\\
\end{matrix}
\end{equation*}

\noi The one-loop calculation of the fermion
self-energies leads to (the first
term corresponding to the neutral current intervention ; the second, to
the charged current intervention -- see figure 1): 
\begin{equation}
\begin{matrix}
Z^{u}_{L}
=
1 - \frac{\epsilon}{2} [\Gamma_{u} \Gamma_{u}^{\dag} + \Gamma_{d} \Gamma_{d}^{\dag} ] & & &
Z^{d}_{L}
=
1 - \frac{\epsilon}{2} [\Gamma_{d} \Gamma_{d}^{\dag} + \Gamma_{u} \Gamma_{u}^{\dag} ] \\
Z^{u}_{R}
=
1 - \frac{\epsilon}{2} [\Gamma_{u} \Gamma_{u}^{\dag} + \Gamma_{u} \Gamma_{u}^{\dag} ] & & &
Z^{d}_{R}
=
1 - \frac{\epsilon}{2} [\Gamma_{d} \Gamma_{d}^{\dag} + \Gamma_{d} \Gamma_{d}^{\dag} ]\\
Z_{M_{u}}
=
1 - \epsilon [ 0 + \Gamma_{d} \Gamma_{d}^{\dag}] & & &
Z_{M_{d}}
=
1 - \epsilon [ 0 + \Gamma_{u} \Gamma_{u}^{\dag}]
\end{matrix}
\label{conren}
\end{equation}

\noi with $\epsilon = \left (\frac{1}{4\pi^{2}}\ln
  \frac{\Lambda^{2}}{\mu^{2}}\right )$ where
$\Lambda$ is a cut-off and $\mu$ an arbitrary energy scale (one checks that
$Z^{u}_{L} = Z^{d}_{L}$ as expected). Those results
are true only up to a term proportional to the identity in the flavour
space, which would take into account the electromagnetic, weak
transversal and strong contributions, as well as the tadpole ones. 
But since this term would factor out in the final result, which is
supposed to 
involve exclusively mixing parameters and {\it up}- or {\it down}-type mass
ratios, we chose not to write it down. The finite parts of the
diagrams are omitted. 

From (\ref{conren}), one derives the corrections to the mass matrices in
the weak base
\begin{align*}
M'_{u} & = M_{u} 
+ \epsilon \;[ M_{d}M_{d}^{\dag}M_{u} - M_{u}M_{u}^{\dag} M_{u} ]
\nonumber\\
M'_{d} & = M_{d} 
+ \epsilon \;[ M_{u}M_{u}^{\dag}M_{d} - M_{d}M_{d}^{\dag} M_{d} ]
\end{align*}

\noi and in the physical base
\begin{align*}
U_{L}^{\dag}M'_{u}U_{R} & = D_{u} 
+ \epsilon \;[K D_{d}^{2}K^{\dag}D_{u} - D_{u}^{3} ]
\nonumber\\
V_{L}^{\dag}M'_{d}V_{R} & = D_{d} 
+ \epsilon \;[K^{\dag} D_{u}^{2}KD_{d} - D_{d}^{3} ]
\end{align*}

\noi where we have absorbed a $\frac{3}{4} \frac{2}{v^{2}}$ factor in $\epsilon$,
$v$ being the scalar VEV, 
and where $D_{u}$ and $D_{d}$ are
the tree-level diagonal mass matrices while $K$ is the tree-level
Cabibbo-Kobayashi-Maskawa matrix.  Let us repeat that those expressions
do not include the finite parts of the radiative correction, and that
they account for the sole (neutral and charged) scalar exchanges
in the fermion self-energies. 

One can rewrite those last expressions as follows
\begin{align*}
M''_{u}& = D_{u} 
+ \epsilon _{u}
\nonumber\\
M''_{d}& = D_{d} 
+ \epsilon _{d}
\end{align*}

\noi and proceed to the diagonalization of $M''_{u}$ and $M''_{d}$,
i.e. 
\begin{align*}
U_{L}^{\prime \dag}M''_{u}U'_{R} & = D'_{u} 
\nonumber\\
V_{L}^{\prime \dag}M''_{d}V'_{R} & = D'_{d} 
\end{align*}

\noi where, in the two-fermion generations case, 
\begin{equation*}
D'_{u}=
\begin{pmatrix}
m_{u}+\epsilon_{u\, 11} &
\\
&
m_{c}+\epsilon_{u\, 22}
\end{pmatrix}
\qquad
U'_{L}=
\begin{pmatrix}
1 & \theta _{u} \\
-\theta _{u} & 1
\end{pmatrix}
\end{equation*}
\begin{equation*}
D'_{d}=
\begin{pmatrix}
m_{d}+\epsilon_{d\, 11} &
\\
&
m_{s}+\epsilon_{d\, 22}
\end{pmatrix}
\qquad
V'_{L}=
\begin{pmatrix}
1 & \theta _{d}\\
-\theta _{d} & 1
\end{pmatrix}
\end{equation*}

\noi with 
\begin{equation*}
\theta _{u}= \frac 
{m_{u}\epsilon_{u\, 21} + m_{c}\epsilon_{u\, 12}} 
{m_{c}^{2}-m_{u}^{2}}
\qquad\mbox{and}\qquad
\theta _{d}= \frac 
{m_{d}\epsilon_{u\, 21} + m_{s}\epsilon_{u\, 12}} 
{m_{s}^{2}-m_{d}^{2}}\quad .
\end{equation*}

\noi The one-loop mixing matrix is defined by 
\begin{equation*}
K' = U^{\prime \dagger}_{L} K V'_{L}
\end{equation*}

\noi so that the one-loop mixing angle reads
\begin{equation*}
\theta' _{C}= \theta _{C}- \theta _{u} + \theta _{d}\quad .
\label{varckm}
\end{equation*}

\noi Inserting the value of $\epsilon_{u}$ and $\epsilon_{d}$ in $\theta'_{C}$,
$D'_{u}$ and $D'_{d}$, leads to equations (\ref{varpara}).

%
%

\bibliographystyle{unsrt}
\bibliography{Bibli}

\vspace{10cm}

\hspace{-8mm}
\includegraphics[width=127mm,keepaspectratio]{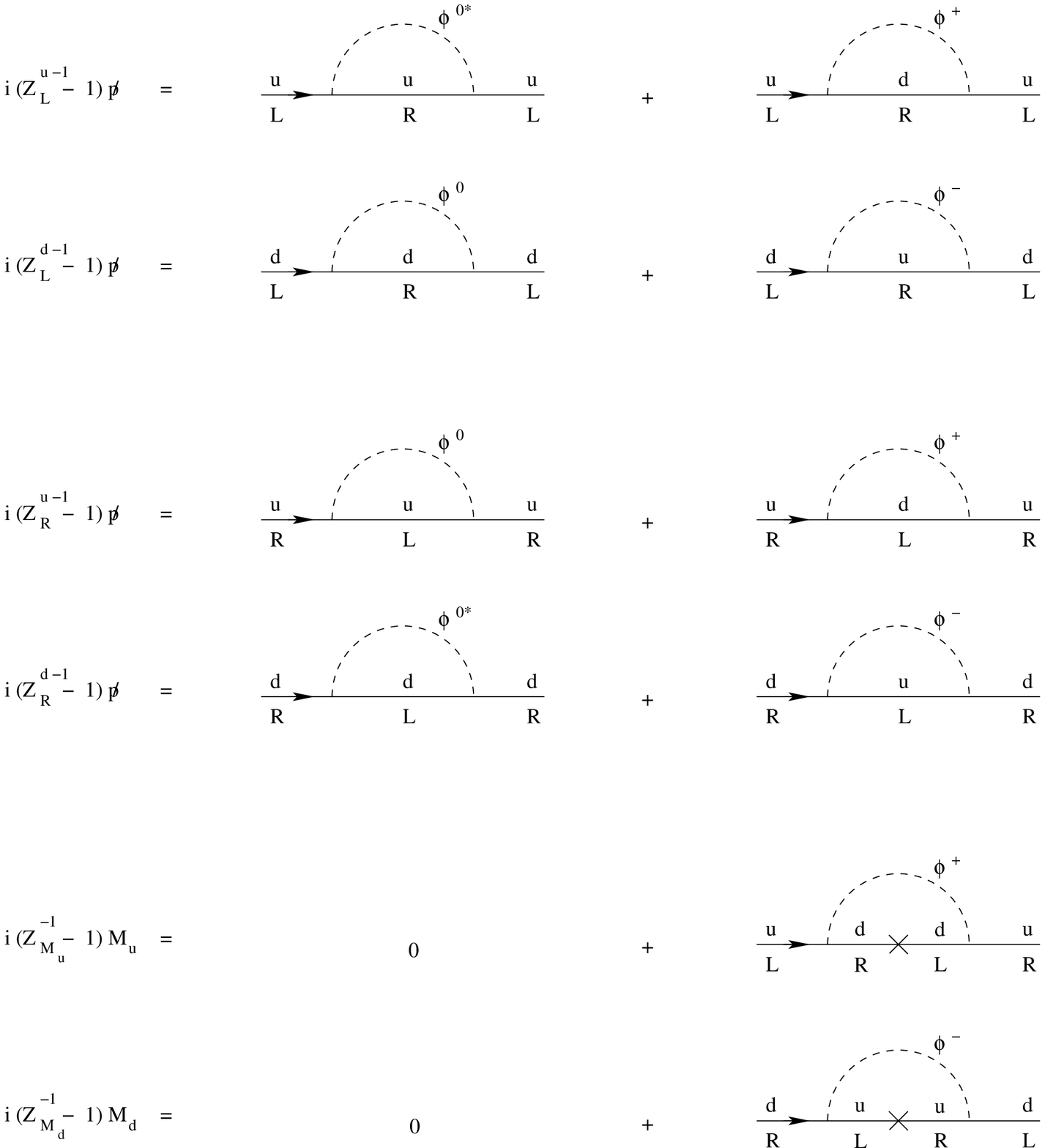}

\vspace{12mm}

\noi Fig. 1: relevant divergent diagrams involved in the calculation
of the renormalization constants.

\end{document}